\documentclass[aps,preprint,superscriptaddress,floatfix]{revtex4}
\usepackage{graphicx}  
\usepackage{dcolumn}   
\usepackage{bm}        
\usepackage{amssymb}   
\usepackage{chapterbib}
\begin{document}

\title{Hybrid modes in a single thermally excited asymmetric dimer antenna}
\author{Loubnan Abou-Hamdan}

   \affiliation{Institut Langevin, ESPCI Paris, PSL University, CNRS, 1 rue Jussieu, F-75005 Paris, France 
}%
                             
     \affiliation{DOTA, ONERA, Université Paris-Saclay, F-91123 Palaiseau, France 
}%
    \author{Claire Li}                               
             \affiliation{Institut Langevin, ESPCI Paris, PSL University, CNRS, 1 rue Jussieu, F-75005 Paris, France 
}%

     \affiliation{DOTA, ONERA, Université Paris-Saclay, F-91123 Palaiseau, France 
}%
     
  \author{Riad Haidar}
  
   \affiliation{DOTA, ONERA, Université Paris-Saclay, F-91123 Palaiseau, France 
}%
  
\author{Valentina Krachmalnicoff}

 \affiliation{Institut Langevin, ESPCI Paris, PSL University, CNRS, 1 rue Jussieu, F-75005 Paris, France 
}%

\author{Patrick Bouchon}
 \email{Corresponding author: patrick.bouchon@onera.fr}
 \affiliation{DOTA, ONERA, Université Paris-Saclay, F-91123 Palaiseau, France 
}%

\author{Yannick De Wilde}   
     \email{Corresponding author: yannick.dewilde@espci.fr}
      \affiliation{Institut Langevin, ESPCI Paris, PSL University, CNRS, 1 rue Jussieu, F-75005 Paris, France 
}%


\begin{abstract}
The study of hybrid modes in a single dimer of neighboring antennas is an essential step to optimize
the far-field electromagnetic (EM) response of large-scale metasurfaces or any complex antenna structure made up of subwavelength building blocks. Here  we  present  far-field infrared  spatial  modulation  spectroscopy (IR-SMS)  measurements of a single thermally excited asymmetric dimer of square metal-insulator-metal (MIM) antennas  separated  by  a  nanometric  gap. Through thermal fluctuations, all the EM modes of the antennas are excited and hybrid bonding and anti-bonding modes can be observed simultaneously. We study the latter within a plasmon hybridization model, and analyse their effect on the far-field response.
\end{abstract}

\maketitle

\begin{figure}[t]
\centering
\includegraphics[height=0.42\textheight]{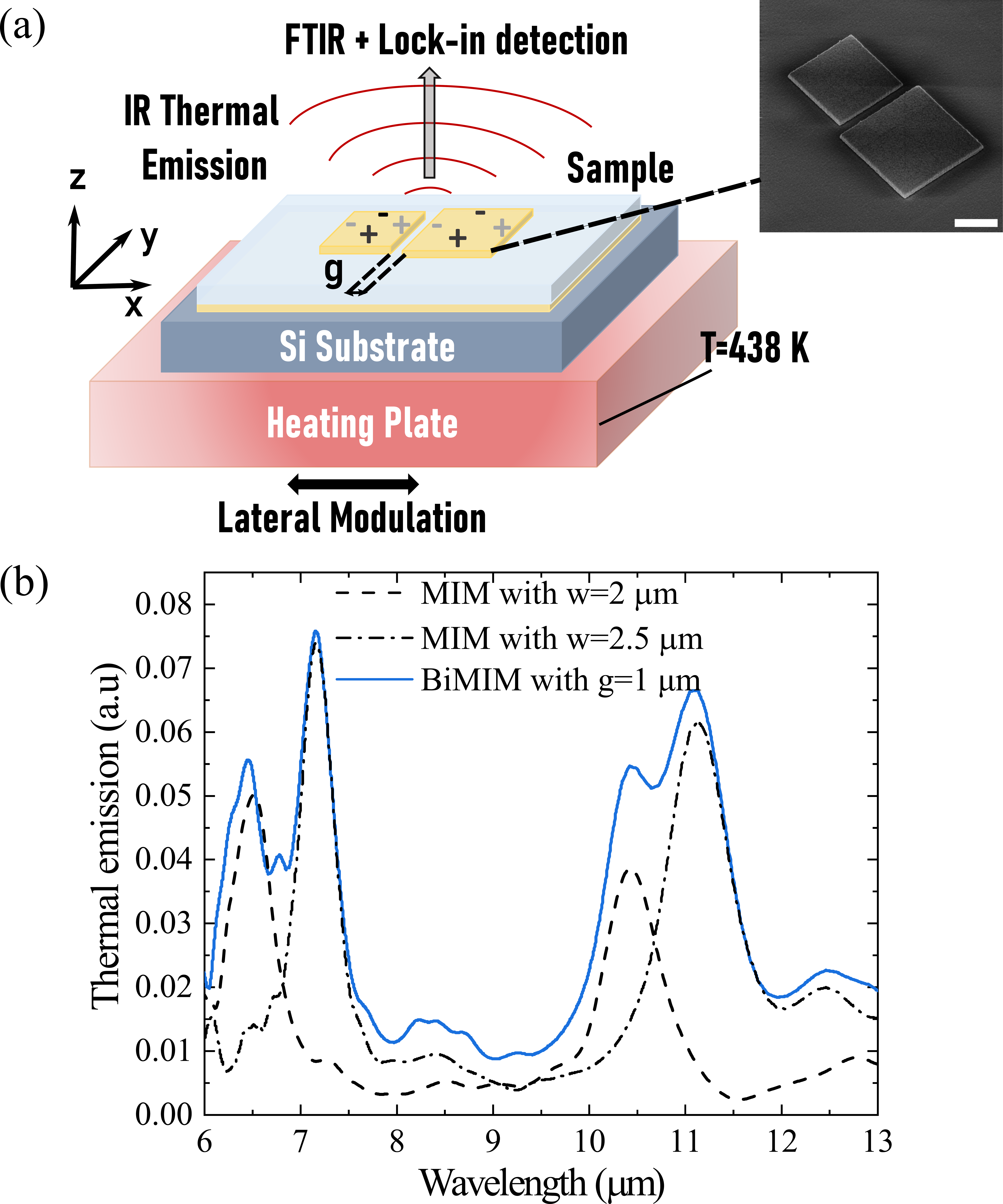}
\caption{(a) Schematic illustration of the BiMIM geometry and IR-SMS technique. The sample consists of a transparent Si substrate on top of which a 200 nm thick layer of gold is evaporated, then a 300 nm spacer layer of SiO$_{2}$ is deposited, and a final top layer consisting of two subwavelength gold square patches of thickness 100 nm, and sides $w_{1}=2$ $\mu$m and $w_{2}=2.5$ $\mu$m, separated by a gap $g$. An SEM image of the investigated sample is shown in the top right corner. The scale bar is $1 \mu$m. (b) Measured IR-SMS far-field thermal emission spectrum of a BiMIM with a 1~$\mu$m gap compared to that of its constituent MIM antennas. The antennas resonate near (6.5~$\mu$m, 10.5~$\mu$m) and (7.2~$\mu$m, 11.2~$\mu$m), for $w=2$~$\mu$m and $2.5$~$\mu$m, respectively. See Supplementary document, section~1 for measurement details.}
\label{fig1}
\end{figure}

Infrared (IR) metamaterials made up of plasmonic resonators have been used in a wide range of applications, such as radiative cooling \cite{raman2014passive, yu2019large}, photodetection \cite{landy2008perfect,palaferri2018room,bouchon2012wideband,fix2017nanostructured,nga2014antenna}, and solar cell design \cite{kelzenberg2010enhanced}. The building blocks of these materials are subwavelength structures that exhibit many interesting antenna effects, namely, directional thermal emission, spectral selectivity, as well as high field confinement. In this regard, MIM resonators have been studied in a variety of configurations, ranging from nanoribbons \cite{todorov2010optical,koechlin2011total} to coated spherical nanoparticles \cite{ yu2019large}, and have demonstrated such capabilities while also being spectrally tunable and angularly independent \cite{todorov2010optical,koechlin2011total,hao2010high}. So far, these studies have mainly considered periodic arrays of antennas. However, diffraction orders as well as coupling effects between neighboring antennas take place, which affects the overall EM response of such arrays \cite{chevalier2015absorbing}.  In addition, when two plasmonic resonators are in near-field interaction, the dimer structure may exhibit hybrid plasmonic resonances \cite{rechberger2003optical,nordlander2004plasmon,halas2011plasmons,gunnarsson2005confined,jain2007universal,brown2010heterodimers, bellido2017plasmonic}. Diffractive coupling between neighboring dimers placed in an array has also been shown to alter optical behavior  \cite{zilio2015hybridization}. The thermal emission of an isolated MIM antenna has been characterized in the mid-IR and has revealed interesting resonant behavior~\cite{li2018near}. Therefore, understanding the effect of near-field interaction within an isolated asymmetric dimer on its far-field thermal emission properties is crucial to guide future efforts of optimizing the response of antennas made of the assembly of sub-wavelength building blocks separated by tiny gaps, with benefit for various applications, such as IR photodetection \cite{palaferri2018room} or multispectral biosensing \cite{adato2009ultra,wu2012fano}. \\
In this letter, we demonstrate simultaneous probing of the bonding and anti-bonding modes of an isolated dimer of MIM antennas by thermal excitation. This striking result, which is confirmed by finite difference time domain (FDTD) simulations and polarization analysis, shows that various coupled modes of a single nano-antenna can be simultaneously excited by thermal fluctuations, an essentially incoherent process arising from fluctuating thermal currents. \\
The investigated samples consist of a homogeneous substrate supporting two coupled subwavelength gold square patch MIM antennas of different width with a silica (SiO$_{2}$) spacer layer, separated by a nanometric gap $g$, which we refer to as a BiMIM structure (see Fig.~\ref{fig1}(a) for sample geometry).  The far-field thermal radiation spectrum of the coupled MIM antennas is extracted from the overwhelming background thermal radiation via IR-SMS \cite{li2018near} in the spectral range between 6-13~$\mu$m. As sketched in Fig.~\ref{fig1}~(a), the temperature of the BiMIM and the substrate is raised to  438~K by means of a hot sample stage. A lateral modulation of the sample at frequency $\Omega$ is applied, producing a small oscillatory contribution on the thermal radiation captured by an IR detector due to a light fall-off effect. Provided that the substrate is homogeneous, it coincides with the sole BiMIM contribution  which is then extracted by lock-in demodulation, allowing background-free spectroscopic analysis with a Fourier transform IR (FTIR) spectrometer. 

A single MIM antenna with a patch of width $w$ acts as a Fabry-Perot resonator for the gap plasmons  trapped below the patch \cite{collin2007waveguiding,todorov2010optical,yang2012ultrasmall}. By thermally exciting the MIM antennas, all their EM modes are excited in the spectral range considered and are populated according to Bose-Einstein statistics \cite{de2006thermal,babuty2013blackbody,carminati2015electromagnetic}. The resonance wavelength of the modes can be roughly estimated by the phase matching condition $\lambda = 2n_{eff}w$, where $n_{eff}$ is the effective index of refraction of the MIM mode in the insulator layer \cite{collin2007waveguiding,yang2012ultrasmall}. Fig. \ref{fig1}(b) shows the measured thermal emission spectrum of a BiMIM structure with a $1\,\mu$m gap (blue curve). As a reference, we also show measured spectra of the single MIM antennas making up the structure (respective widths $w=2$ $\mu$m and $2.5$~$\mu$m, dashed and dash-dotted curves) each consisting of two resonance peaks, which are a signature of the excitation of the MIM fundamental mode at two different wavelengths. The peak at the largest wavelength results from the peculiar dispersion of the silica spacer layer \cite{li2018near} and will not be discussed further here as it occurs in a region where the cavity is very lossy (see supplementary document, section 2 for more details). We will thus focus the remainder of our discussion on spectral features below 8.5~$\mu$m.
As can be seen from Fig.~\ref{fig1}(b), for such a large gap, the thermal emission spectrum
of the BiMIM structure simply overlaps with that of the two
individual antennas, meaning that the two MIM antennas behave independently. 
\begin{figure}[t]
\centering
\includegraphics[width=\linewidth]{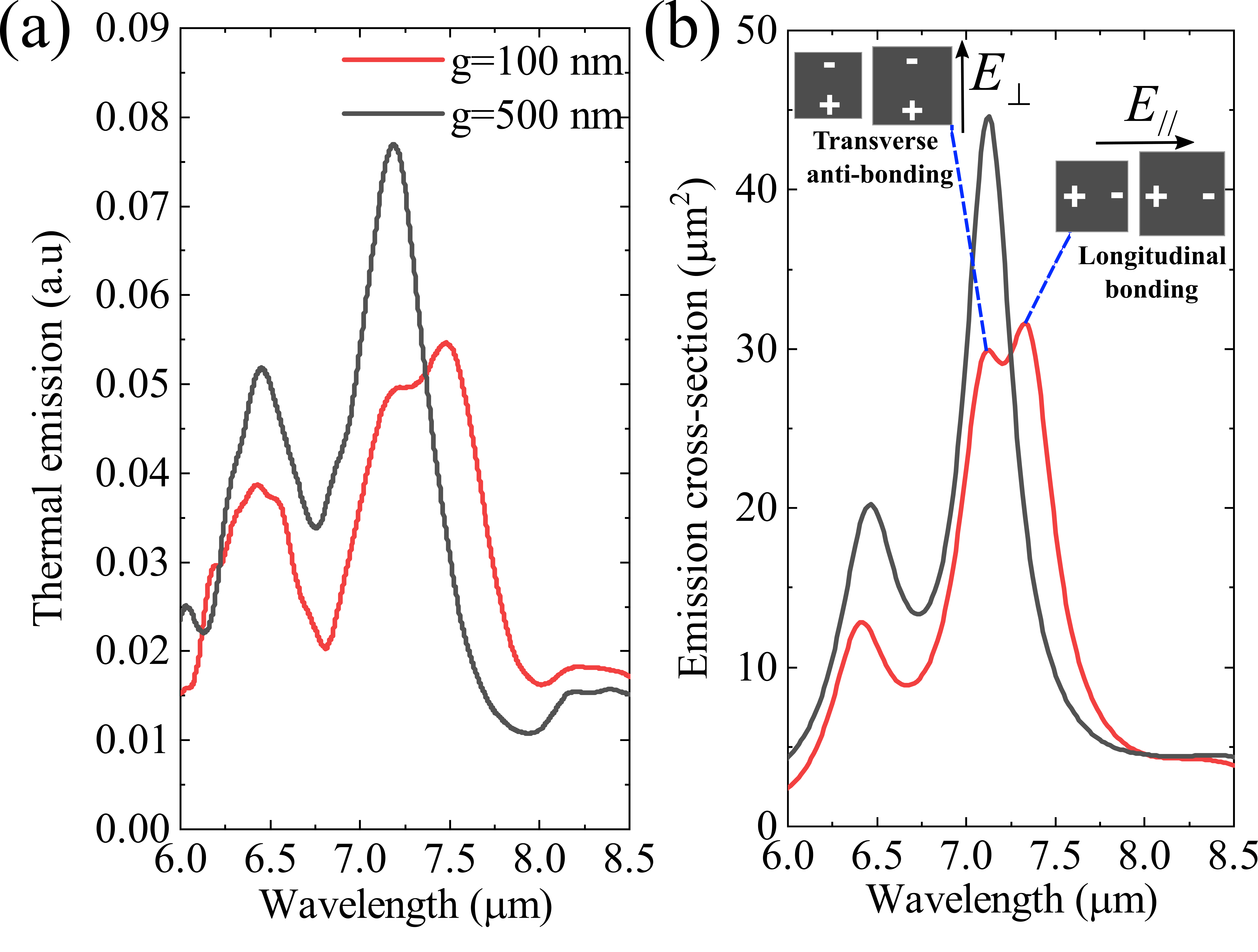}
\caption{(a) Measured thermal emission spectra of the BiMIM structure for different gap size g. (b) FDTD calculations of emission cross-sections for normally incident, unpolarized illumination, corresponding to the measurements (see Supplementary document, section 2 for calculation details). The inset shows a sketch of the dimer geometry showing two of the hybrid modes of the BiMIM, for electric field polarizations parallel and perpendicular to the dimer axis. The positive and negative signs indicate the surface
charge distribution.  }
\label{fig2}
\end{figure}
However, when the MIM antennas are separated by a smaller gap, their two gold patches form a plasmonic dimer pair that exhibits bonding and anti-bonding hybrid modes \cite{nordlander2004plasmon}. If an external illumination is used, the latter modes can be selectively excited by varying the polarization of the EM excitation with respect to the dimer common axis (see Fig.\ref{fig2}(b) inset). These hybrid modes can be studied using a plasmon hybridization model \cite{prodan2004plasmon,prodan2003hybridization,nordlander2004plasmon}, in which the conduction electrons in the metal are modeled as a charged incompressable liquid, whose deformations lead to the formation of a surface charge.\\
In our experiment, a thermal excitation of the BiMIM structure is instead produced in situ by raising the overall temperature of the sample, inducing a superposition between the longitudinal and the transverse bonding and anti-bonding modes, which modifies the observed far-field response. In particular, when  the gap  size  is  reduced  to  100 nm (red curve in Fig. \ref{fig2}) a splitting  begins  to form  in  the resonance  peak  between  7  and  8 $\mu$m, while an overall red-shift in the peaks is also observed. It has been shown that the bonding hybrid modes lead to a red-shift in the spectrum, while the anti-bonding modes result in a slight blue-shift \cite{rechberger2003optical,gluodenis2002effect,gunnarsson2005confined,jain2007universal}.
\begin{figure}[t]
\centering\includegraphics[height=0.42\textheight]{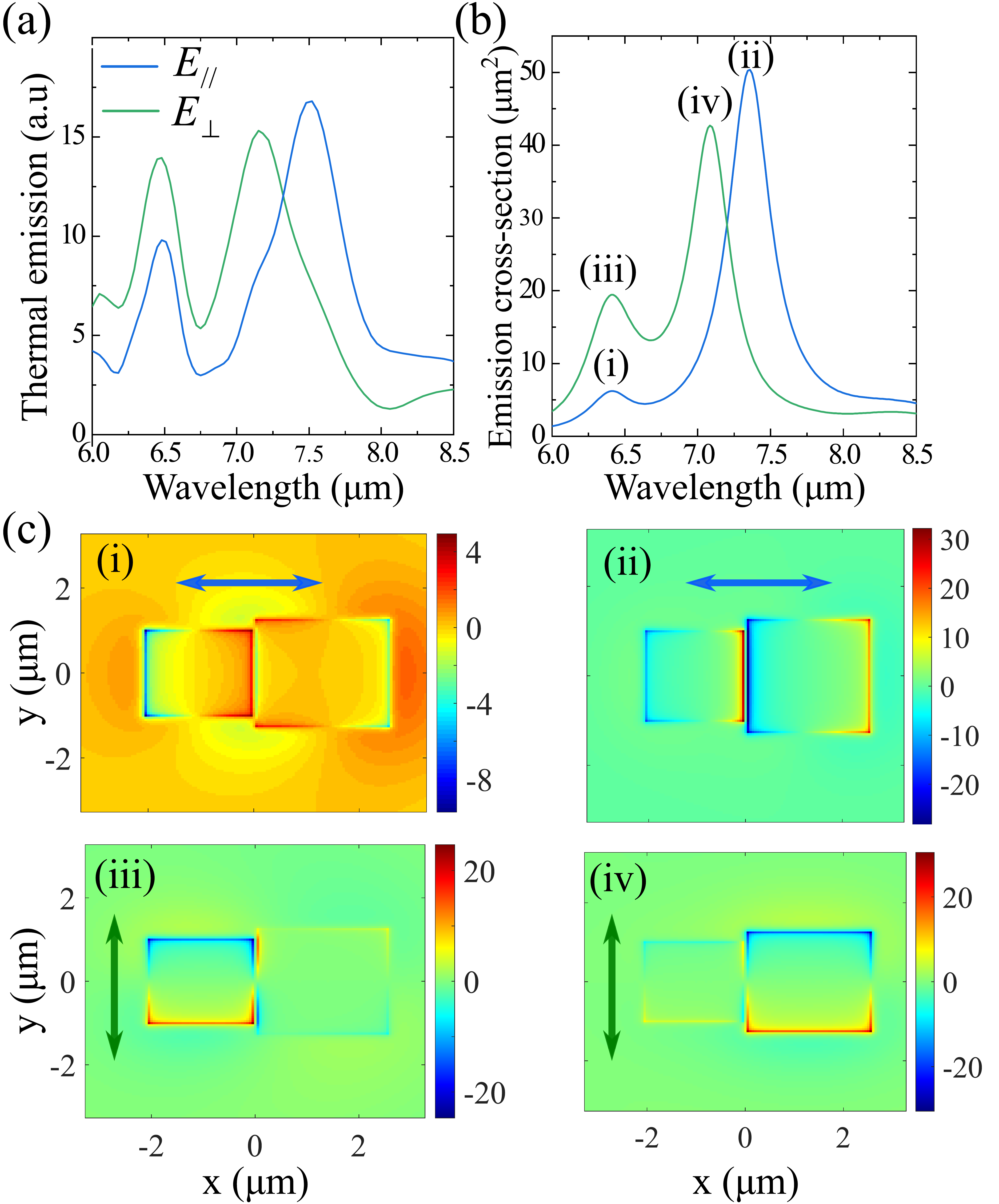}
\caption{(a) Measured thermal emission and (b) calculated emission cross-section of the BiMIM structure with $g=100$ nm for polarized thermal emission. (c) Plots of the z component $E_{z}$ (in units of V/m) of the electric field in the xy-plane at the top of the square metal patches, at the resonance peaks indicated by the markers (i), (ii), (iii), and (iv) in panel (b). The arrows indicate the electric field polarization in each case.}
\label{fig3}
\end{figure} 
The observed splitting is thus, a hallmark of simultaneous excitation of the longitudinal bonding and transverse anti-bonding modes due to thermal fluctuations. FDTD calculations of the emission cross-section of the BiMIM structure (Fig. \ref{fig2}(b)) confirm this effect. For a larger gap size ($g=500$ nm, black curve in Fig. \ref{fig2}) the splitting and red-shift disappear. \\
The thermal radiation emitted by the BiMIM structure is collected by a reflective objective ($NA=0.5$), which collects radiation from a solid angle between 10 and 30$^{o}$. By reciprocity this means that the BiMIM’s hybrid modes can be excited by an external source at non-normal incidence. The agreement of our off-axis measurements with simulations at normal incidence suggests that the emission spectra of the hybrid modes of the considered BiMIM structure are angularly independent up to at least 30$^{o}$. \\
As shown below, it is possible to distinguish between the various hybrid modes of the BiMIM antennas, that are thermally excited, by performing polarized thermal emission measurements. This is done by placing a wire grid polarizer before the detector in order to select the $E_{//}$ and $E_{\perp}$ polarizations. The experimental results and  the corresponding cross-section calculations are shown in Fig.~\ref{fig3}~(a) and (b) respectively. For the parallel polarization, FDTD simulations show that the longitudinal anti-bonding and bonding modes are excited at the resonance positions (i) and (ii), respectively, in Fig. \ref{fig3}(b). At these resonance positions, charges accumulate on the left and right edges of the  square patches. This produces a strong enhancement of the $z$ component $|E_{z}|$ of the electric field  at the edges of the squares, as shown in Fig. \ref{fig3}(c) (i-ii). The opposite charges, found at the edges near the gap, lead to the observed red-shift in the longitudinal bonding mode (position (ii) in Fig. \ref{fig3}(b)), due to attraction. On the other hand, charges of same sign are found at the edges near the gap for the longitudinal anti-bonding mode (Fig. \ref{fig3}(c) (i)). This latter mode appears as a small peak at position (i) in Fig. \ref{fig3}(b).
\begin{figure}[t]
\centering\includegraphics[height=0.43\textheight]{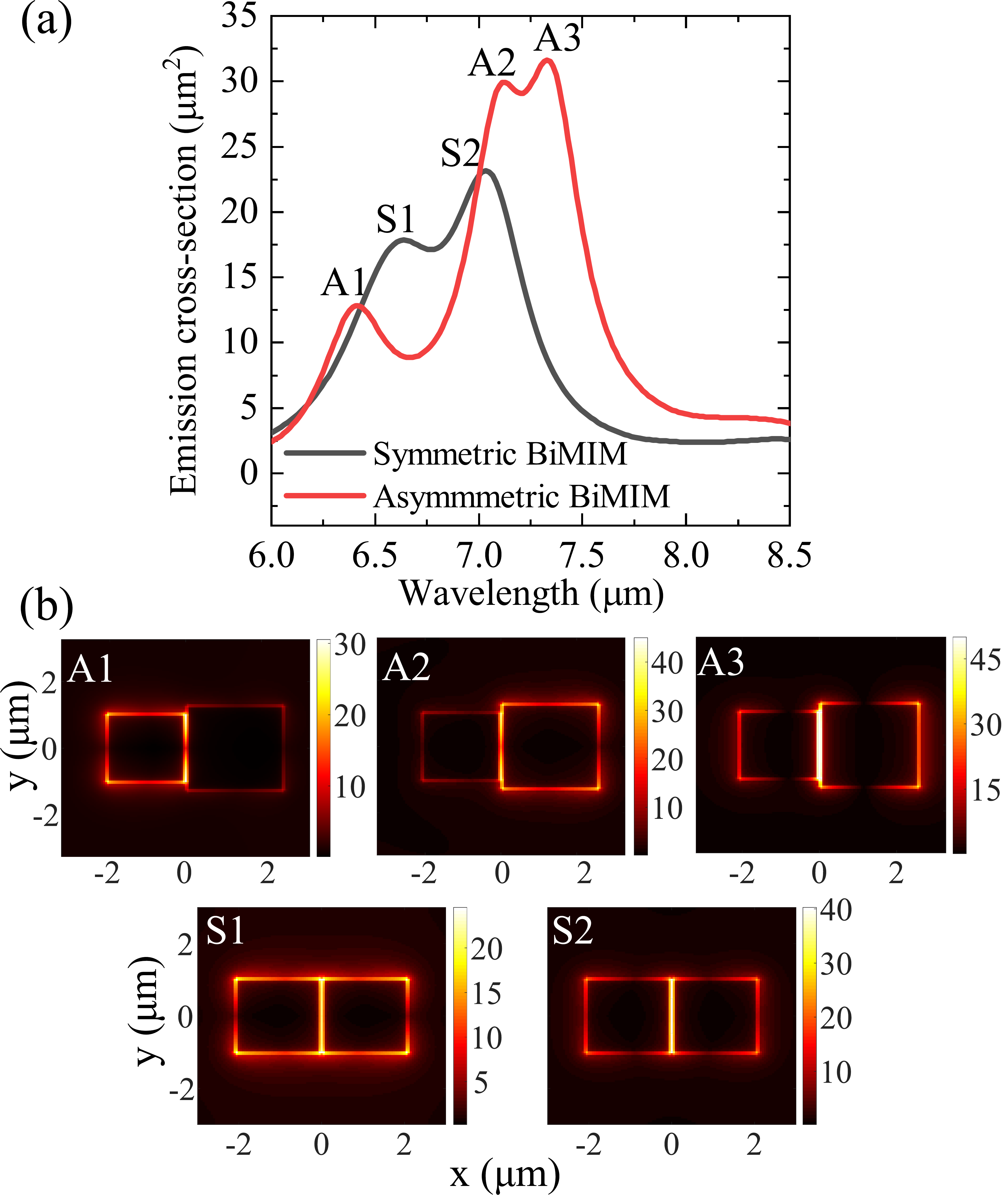}
\caption{(a) Simulated emission cross-section for unpolarized light for the considered BiMIM structure with $g=100$~nm  compared with that of a symmetric BiMIM whose patches have widths $w_{1}=w_{2}=2$ $\mu$m. (b) Electric field enhancement plots $|E|/|E_{o}|$, where $|E_{o}|$ is the incident field, showing field hot spots in the xy-plane at the top of the metal patches, recorded at the resonance positions in panel~(a). }
\label{fig4}
\end{figure} 
Note that the asymmetry of the  BiMIM structure is essential for the observation of this mode. If the square patches were identical in size the net dipole moment for this mode would be zero, and it would be inaccessible by far-field illumination, as it is a dark mode \cite{prodan2004plasmon}. By reciprocity it would not produce any measurable thermal radiation to the far-field as well.
For the $E_{\perp}$ polarization, charges accumulate at the top and bottom edges of the patches, leading to the weakly interacting transverse bonding and anti-bonding modes at positions (iii) and (iv), respectively (Fig.~\ref{fig3}(c) (iii-iv)). In the measured spectra (Fig.~\ref{fig2}(a)), the thermally induced splitting is thus the result of  the combined response of the longitudinal and transverse bonding and anti-bonding modes.  In contrast, the longitudinal anti-bonding and the transverse bonding modes (i and iii, respectively) are degenerate and do not produce a splitting in the unpolarized thermal emission spectrum.\\
If a symmetric BiMIM structure is considered (where both MIM antennas of the structure resonate near 6.5 $\mu$m),  Fig.~\ref{fig4} (black simulated curve), the spectral overlap of the resonances near 6.5~$\mu$m does not allow for the excitation of the dark mode, while a splitting due to the longitudinal bonding and transverse anti-bonding modes still occurs–- as indicated by the splitting between 6 and $7.5\,\mu$m. 
An interesting effect is also observed in the considered BiMIM structure, when examining the simulated electric field enhancement plots, for unpolarized emission, at the resonance positions A1 and A2 (Fig.~\ref{fig4} (b)). The plots reveal emission hot spots from the BiMIM structure, showing that the two MIM antennas emit independently at these resonance positions, even though they are effectively in near-field interaction. On the other hand, at the resonance peak A3 the two MIM antennas act as a coupled system, and a larger field enhancement is found in the gap (Fig.~\ref{fig4} (b), A3). The observed behavior at positions A1 and A2 is a direct result of the nature of the weakly interacting transverse hybrid modes, and the asymmetry of the BiMIM structure under study. In fact, since the resonance wavelength of a MIM antenna scales with its width, the size difference between the two MIM antennas allows for independent emission at two different wavelengths. To illustrate this point, we also show field enhancement plots at the resonance positions S1 and S2 for the symmetric BiMIM case (Fig~\ref{fig4}. (b) bottom two panels). In this case, the two MIM antennas cannot emit independently, and emission hot spots are either found at the edges of the two gold patches (position S1) or at the gap (position S2).\\
The equivalence between emission and absorption suggests that if an external electric field is incident on the considered asymmetric BiMIM structure, it will be routed to, and resonantly absorbed by each MIM antenna at the resonance positions A1 and A2. Such an attribute could be beneficial for IR photodetection applications \cite{koechlin2011total}.\\
In summary, we have shown thermally induced simultaneous probing of the various hybrid modes of a single asymmetric BiMIM structure, and measured its effect on the far-field response using IR-SMS. In contrast to other symmetric systems the considered asymmetric BiMIM structure was found to allow direct far-field excitation of the dark anti-bonding mode. Further, simulated electric field enhancement plots revealed that due to the nature of the transverse hybrid modes, and the asymmetry of the system considered, the two MIM antennas making up the structure can emit (equivalently absorb) radiation independently. This feature, along with the fact that the absorbed radiation is mainly concentrated in the insulator layer in a MIM antenna, could be of use for realizing IR photodetectors with improved responsivity at room temperature. The absorbed radiation may be harnessed by replacing the insulating SiO$_{2}$ layer considered here by a semi-conductor like HgCdTe \cite{le2009plasmon,koechlin2011total} or a quantum well material \cite{palaferri2018room}.  Finally, our study may guide future efforts to optimize near-field interactions in nano-antenna arrays for applications such as bio-sensing for the detection and characterization of bio-molecules. 
\section*{Acknowledgments}
The authors would like to thank Christophe Dupuis for the SEM image. This work received financial support from LABEX WIFI under references ANR-10-LABX-24 and ANR-10-IDEX-0001-02 PSL*, and ANR Project CarISOVERRE ANR-16-CE09-0012.
\bibliography{references}

\end{document}


\author{Loubnan Abou-Hamdan}

   \affiliation{Institut Langevin, ESPCI Paris, PSL University, CNRS, 1 rue Jussieu, F-75005 Paris, France 
}%
                             
     \affiliation{DOTA, ONERA, Université Paris-Saclay, F-91123 Palaiseau, France 
}%
    \author{Claire Li}                               
             \affiliation{Institut Langevin, ESPCI Paris, PSL University, CNRS, 1 rue Jussieu, F-75005 Paris, France 
}%

     \affiliation{DOTA, ONERA, Université Paris-Saclay, F-91123 Palaiseau, France 
}%
     
  \author{Riad Haidar}
  
   \affiliation{DOTA, ONERA, Université Paris-Saclay, F-91123 Palaiseau, France 
}%
  
\author{Valentina Krachmalnicoff}

 \affiliation{Institut Langevin, ESPCI Paris, PSL University, CNRS, 1 rue Jussieu, F-75005 Paris, France 
}%

\author{Patrick Bouchon}
 \email{Corresponding author: patrick.bouchon@onera.fr}
 \affiliation{DOTA, ONERA, Université Paris-Saclay, F-91123 Palaiseau, France 
}%

\author{Yannick De Wilde}   
     \email{Corresponding author: yannick.dewilde@espci.fr}
      \affiliation{Institut Langevin, ESPCI Paris, PSL University, CNRS, 1 rue Jussieu, F-75005 Paris, France 
}%


\title{Supplementary information}

\maketitle

\section{Experimental details}

\begin{figure}[htb]
\centering\includegraphics[width=\linewidth]{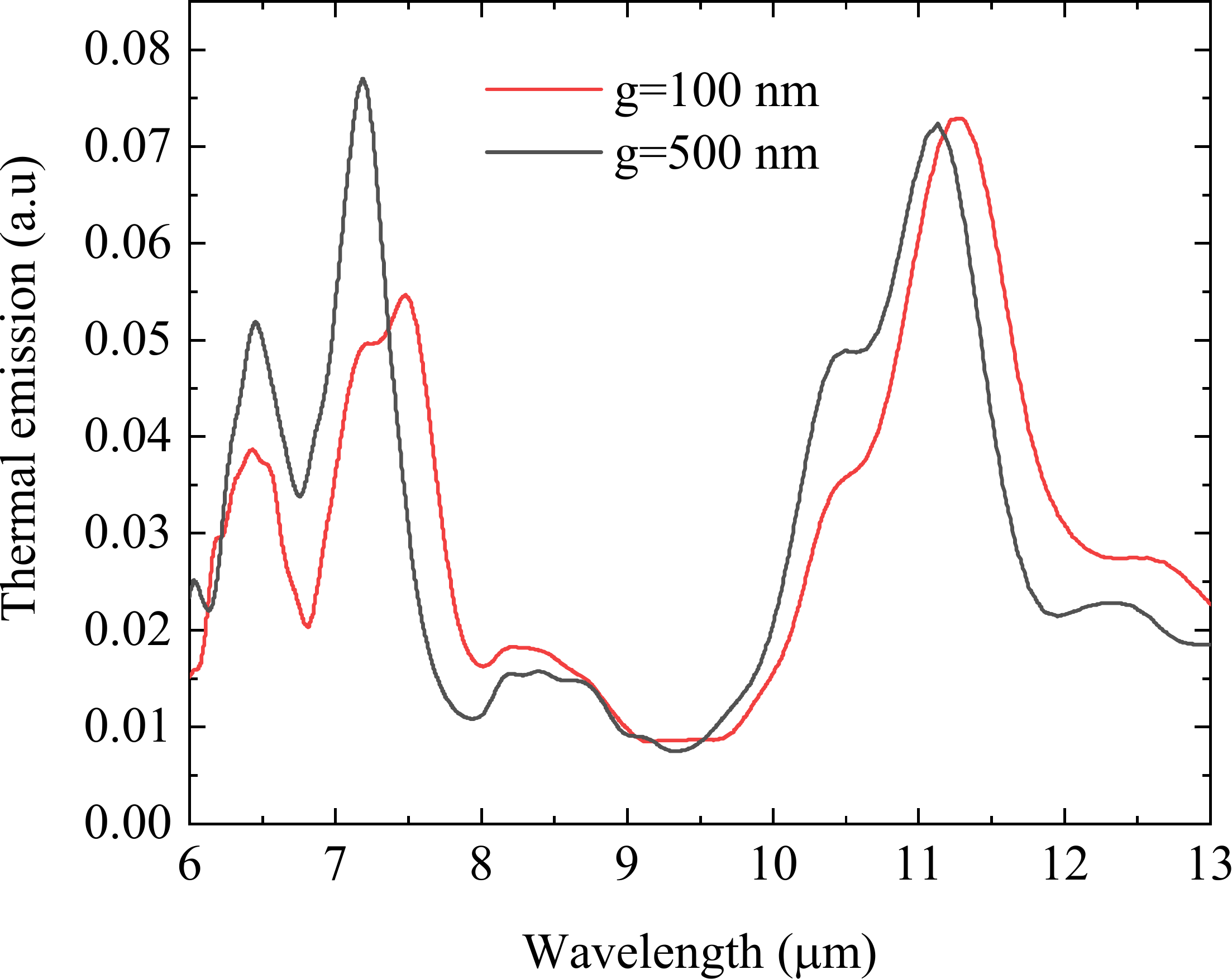}
\caption{Measured thermal emission spectra of the BiMIM structure for different gap sizes $g$ (presented in Fig.~2 (a) of the main text), over the spectral range between 6--13~$\mu$m.}
\label{Sfig1}
\end{figure}

\begin{figure}[htb]
\centering\includegraphics[width=\linewidth]{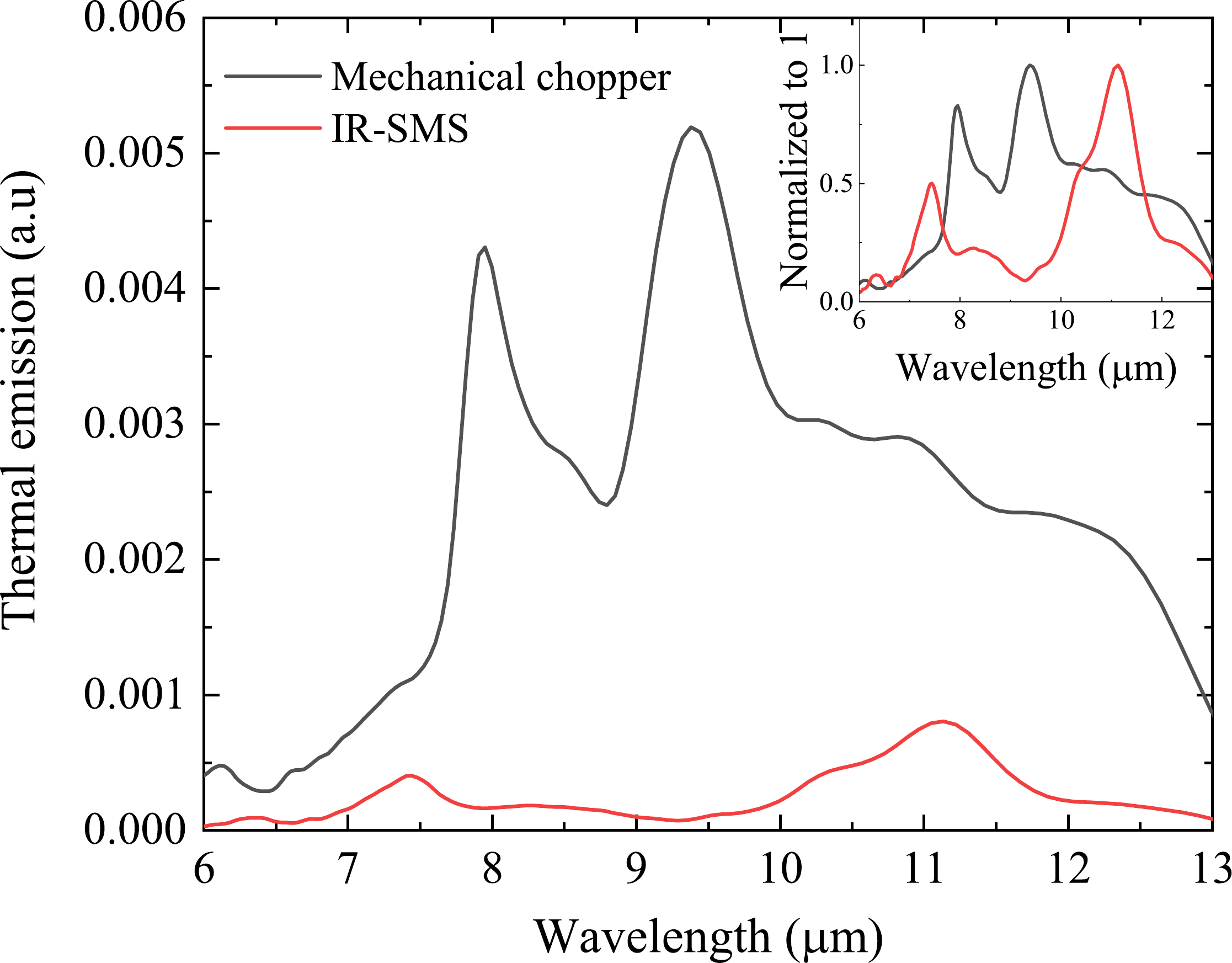}
\caption{Raw thermal emission spectra of the BiMIM structure measured using the IR-SMS technique and a mechanical chopper. The two spectra are scaled according to the sensitivity used in the lock-in detection. The inset shows a comparison of the two when normalized to 1. }
\label{Sfig2}
\end{figure}

The infrared spatial modulation spectroscopy technique (IR-SMS) used to obtain the thermal emission of the isolated BiMIM structure is based on lateral modulation of the sample along with lock-in detection. The full details of the experimental technique have already been outlined elsewhere in Ref. \cite{li2018near}.
The experimental setup is coupled to an FTIR which allows us to obtain the thermal emission spectra shown in Fig. 2(a) of the main text. For completeness, these spectra are again plotted here over the full spectral range of measurement between 6--13~$\mu$m in Fig~\ref{Sfig2}.\\
During the measurements, the sample itself is the source of the thermal emission, which is brought upon by heating it to a uniform temperature. The presented measurements were performed at $T_{sample}=438$~K. The instrumental response of the FTIR was calibrated with blackbody samples at different temperatures using the method outlined in Ref. \cite{revercomb1988radiometric}.  The measured thermal emission spectra, presented in the main text, were normalized to the response of a blackbody at the same temperature to correct for instrumental response. As such, the normalized spectra are directly comparable to the emission cross-section of the BiMIM structure (see ref. \cite{kallel2019thermal} for more details). \\
\begin{figure*}[t]
\centering\includegraphics[width=0.9\linewidth]{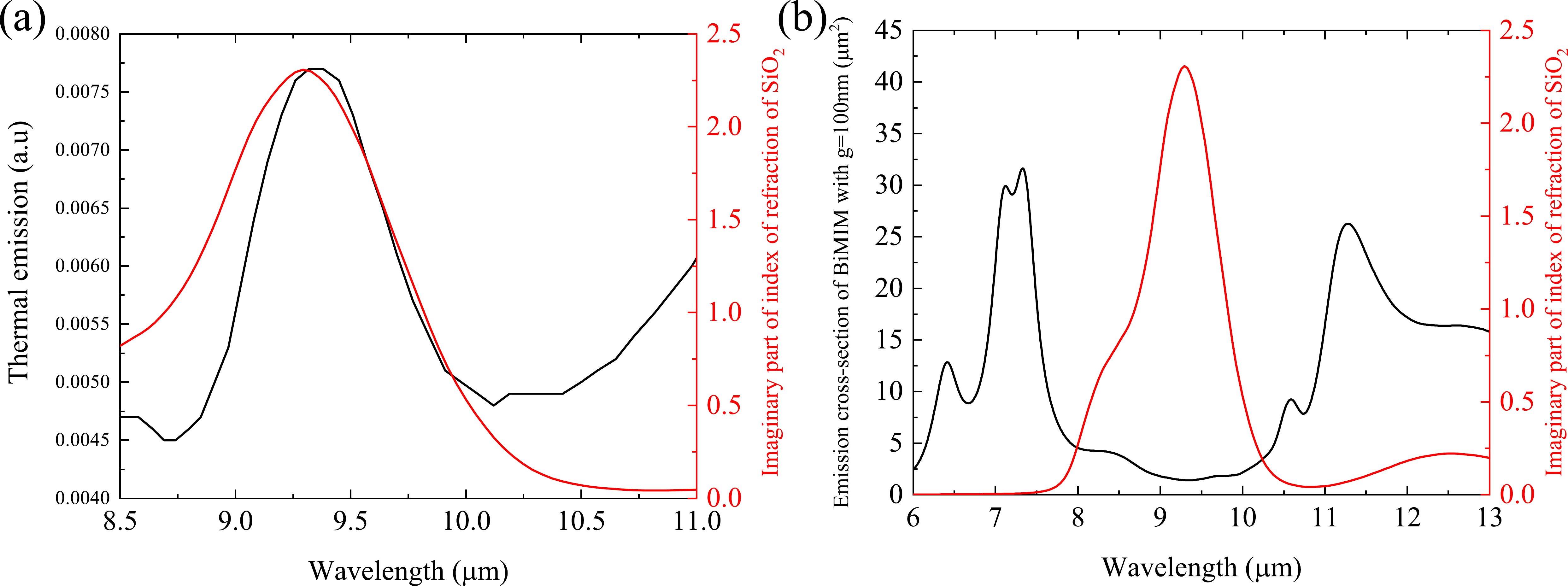}
\caption{(a) Measured thermal emission of the sample substrate, along with the imaginary part of the index of refraction of SiO$_{2}$ (taken from Ref. \cite{kischkat2012mid}) representing the absorption due to the substrate. (b) Simulated emission cross-section of a BiMIM structure for gap size $g=100$ nm (black curve) presented in the main text (red curve in Fig.~2 (b) of the main text), along with the imaginary part of the SiO$_{2}$ index of refraction (red curve). The comparison of the two shows a high absorption due to SiO$_{2}$ in the spectral region between 8 and 13 $\mu$m, which prevents the formation of a splitting in the BiMIM emission spectrum in this range.  }
\label{Sfig3}
\end{figure*}
To illustrate the fact that the IR-SMS technique allows one to obtain a better signal-to-noise ratio and can directly extract the thermal emission of the single BiMIM structure from the huge extended background radiation, we show in Fig. \ref{Sfig2} raw (un-normalized) spectra of the BiMIM structure measured both using the IR-SMS technique and with a mechanical chopper. The two spectra are scaled according to the sensitivity used during the lock-in detection. As clearly indicated in Fig. \ref{Sfig2} the signal coming from the sole BiMIM antenna, which is directly measured by the IR-SMS technique, is much smaller than that of the background emission. On the other hand, the BiMIM emission is completely lost in the overwhelming signal coming from the substrate emission and background thermal radiation when the measurement is performed with a mechanical chopper. The peak near 9.5~$\mu$m in Fig. \ref{Sfig2} (black curve) for instance, is an absorption peak of the SiO$_{2}$ substrate. A comparison of the two spectra when normalized to 1 (Fig. \ref{Sfig2} inset) shows that the resonance peaks of the measured BiMIM structure can be clearly resolved even in the raw untreated spectrum, showing the efficacy of the IR-SMS technique. \\
The polarized thermal emission spectra presented in Fig. 3(a) of the main text, were obtained by adding a BaF$_{2}$ wire grid polarizer-- with maximum transmission in the measured spectral range-- before the detector.

\section{FDTD simulations}

The simulations presented in the main text where performed using FDTD calculations (Lumerical solutions). In order to simulate an isolated BiMIM structure perfectly matched layers (PMLs) were effected for all boundaries. Unpolarized, normally incident, broadband electromagnetic radiation in the 6-13 $\mu$m range was used as the source of excitation. The simulation span was made large enough so that the distance between the edges of the patch antennas of the BiMIM structure and the PMLs was at least $\lambda/4$ in all directions, where $\lambda$ is the maximum wavelength in the considered spectral range. This was done to reduce reflections by the PML boundaries and to improve the convergence of the simulations. The simulation time was also set large enough to insure proper convergence of all simulations. The emission cross-section was obtained by calculating losses due to absorption from the MIM antennas. \\
Palik data \cite{edward1985handbook} were used in the simulations for the dielectric function of the gold layers. On the other hand, dielectric data reported in Ref. \cite{kischkat2012mid} for SiO$_{2}$ thin films were used for the SiO$_{2}$ layer. In order to estimate the difference between the used dielectric data for SiO$_{2}$ and the real one, we measured the thermal emission from the substrate, by collecting thermal radiation from a part of the sample that does not include any gold patches, using a mechanical chopper. The measured thermal emission of the substrate can be compared to the absorption of the SiO$_{2}$ layer, represented by the imaginary part of its index of refraction. As shown in Fig.\ref{Sfig3}~(a) the peak of absorption of the imaginary part of the index of refraction agrees well with the peak of the measured thermal emission.\\
On examining the imaginary part of the index of refraction further, we find that due to the high absorption of SiO$_{2}$ in the spectral range between 8 and 13$\mu$m, a splitting due to the hybrid modes does not occur in this region. Indeed, as illustrated in Fig.\ref{Sfig3}~(b) the imaginary part of the index of refraction is zero in the spectral range where the splitting occurs (below $\sim$7.5 $\mu$m), while it is non-zero for higher wavelengths.

\bibliography{ref_SI}